\documentclass[11pt]{article}
\usepackage[pdftex]{color}
\usepackage{graphicx}
  \usepackage{amsthm,amsfonts}
  \usepackage{amsmath}
\usepackage{slashed}
\usepackage{ulem}
\usepackage{cleveref}
\usepackage{cite}
\newcommand{\bea}   {\begin{eqnarray}}
\newcommand{\eea}   {\end{eqnarray}}

\def\zzg{${\mathbb Z}_2\times{\mathbb Z}_2$-graded }

\begin{document}
\renewcommand{\thefootnote}{\fnsymbol{footnote}}

\thispagestyle{empty}

\title{Beyond the $10$-fold way: $13$ associative \\ $ {\mathbb Z}_2\times{\mathbb Z}_2$-graded  superdivision algebras}

\author{Zhanna Kuznetsova\thanks{{E-mail: {\it zhanna.kuznetsova@ufabc.edu.br}}}\quad and\quad
Francesco Toppan\thanks{{E-mail: {\it toppan@cbpf.br}}}
\\
\\
}
\maketitle

\centerline{$^{\ast}$ {\it UFABC, Av. dos Estados 5001, Bangu,}}\centerline{\it { cep 09210-580, Santo Andr\'e (SP), Brazil. }}
~\\
\centerline{$^{\dag}$
{\it CBPF, Rua Dr. Xavier Sigaud 150, Urca,}}
\centerline{\it{
cep 22290-180, Rio de Janeiro (RJ), Brazil.}}
~\\
\maketitle
\begin{abstract}
The ``$10$-fold way" refers to the combined classification of the $3$ associative division algebras (of real, complex and quaternionic numbers) and of the $7$,  ${\mathbb Z}_2$-graded, superdivision algebras (in a superdivision algebra each homogeneous element is invertible). 
The connection of the $10$-fold way with the periodic table of topological insulators and superconductors is well known. Motivated by the recent interest in ${\mathbb Z}_2\times{\mathbb Z}_2$-graded physics
(classical and quantum invariant models, parastatistics) we classify the associative ${\mathbb Z}_2\times {\mathbb Z}_2$-graded superdivision algebras and show that $13$ inequivalent cases have to be added to the $10$-fold way. Our scheme is based on the ``alphabetic presentation of Clifford algebras", here extended to graded superdivision algebras. The generators are expressed as equal-length words in a $4$-letter alphabet (the letters encode a basis of invertible $2\times 2$ real matrices and in each word the symbol of tensor product is skipped).
The $13$ inequivalent \zzg superdivision algebras are split into real series ($4$ subcases with $4$ generators each), complex series ($5$ subcases with $8$ generators) and quaternionic series ($4$ subcases with $16$ generators).
{{As an application, the connection of ${\mathbb Z}_2\times{\mathbb Z}_2$-graded superdivision algebras with a parafermionic Hamiltonian possessing time-reversal and particle-hole symmetries is presented.}}
\end{abstract}
\vfill

\rightline{CBPF-NF-004/21}

\newpage

\section{Introduction}
Real numbers (${\mathbb R}$), complex numbers (${\mathbb C}$) and quaternions (${\mathbb H}$) are the $3$
associative division algebras over the reals. By allowing a ${\mathbb Z}_2$-grading, $7$ further associative superdivision algebras are encountered (in a graded superdivision algebra, each homogeneous element admits inverse). The total number of $10=3+7$ is known as the ``$10$-fold way", see \cite{bae} for a short introduction on the topic of superdivision algebras.\par
This purely mathematical property has striking connections with physical applications. The $10$-fold way appears in the construction of the so-called ``periodic table of topological insulators and superconductors", which presents features like the $mod~ 8$ Bott's periodicity. The physical significance of the periodic table is discussed in \cite{kit}.
In \cite{{zir},{alzi}} the Cartan's classification of symmetric spaces was related to the $10$ classes of symmetries
of random-matrix theories. The $10$-fold way classification of generic Hamiltonians and topological insulators and superconductors has been discussed in \cite{rsfl} (see also the references therein). On the mathematical side, different implementations of the $10$-fold way have been analyzed in \cite{frmo}; the one which is relevant for our paper (besides the Morita's equivalence of Clifford modules and the graded extension of the Dyson's threefold way \cite{dys}) is the direct relation with  the \cite{wal} classification of graded Brauer groups, i.e. superdivision algebras.\par
Since there is no need to stop at ${\mathbb Z}_2$, the notion of superdivision algebra can be immediately extended to accommodate a ${\mathbb Z}_2\times {\mathbb Z}_2$-grading. The aim of this paper is to classify the inequivalent, associative, \zzg superdivision algebras. We obtain $13$ inequivalent cases. Before presenting the mathematical results we mention that the motivation of this work lies in the recent surge of interest in \zzg physics. We present a succinct state of the art on this topic. \zzg Lie algebras and superalgebras were introduced in \cite{{rw1},{rw2}} (see also \cite{sch}) and investigated by mathematicians ever since.
The attempts to use them for physical applications were rather sporadic. This situation changed when it was pointed out in \cite{{aktt1},{aktt2}} that \zzg Lie superalgebras naturally appear as dynamical symmetries of Partial Differential Equations describing  L\'evy-Leblond nonrelativistic spinors. An ongoing intense activity followed.  A model of \zzg 
invariant quantum mechanics was discussed in \cite{brdu}, while the systematic construction of \zzg classical mechanics 
was introduced in \cite{akt1} and the canonical quantization of the models presented in \cite{akt2}.  Further developments include the construction of two-dimensional models \cite{{bru},{kuto},{bru2}}, graded superconformal quantum mechanics \cite{aad}, superspace \cite{{brdu2}, {kuto},{aido}}, extensions and bosonization of double-graded supersymmetric quantum mechanics \cite{{aad2},{que}}, and so forth. The construction of models with \zzg parabosons was discussed  in \cite{kuto}.
The theoretical detectability of the parafermions implied by the \zzg Lie superalgebras was proved in\cite{top1}, while in \cite{top2} the result was extended to the parabosons implied by \zzg Lie algebras (for previous works on \zzg parastatistics see \cite{{tol},{stvdj}} and the references therein).
All this ongoing activity makes reasonable to expect that the classification of \zzg superdivision algebras would not just be a mathematical curiosity, but could find a way to concrete physical applications. \par
As an efficient tool to classify \zzg superdivision algebras we apply the so-called ``alphabetic presentation of Clifford algebras" introduced in \cite{tove}, which is here extended to superdivision algebras. In this framework the generators are expressed as matrices defined by equal-length words in a $4$-letter alphabet (the letters encode a basis of invertible $2\times 2$ real matrices and in each word the symbol of tensor product is skipped). Each one of the $7$
inequivalent ${\mathbb Z}_2$-graded division algebra admits an alphabetic presentation; it follows from this property that any \zzg superdivision algebra can be alphabetically presented. Within this scheme, spotting the inequivalent classes of \zzg superdivision algebras is reduced to a simple exercise in combinatorics. We obtain $13$ inequivalent cases which are split into $4$ real, $5$ complex and $4$ quaternionic subcases. The number of generators, in each subcase, is respectively $4$, $8$ and $16$. Correspondingly, the faithful representations  are given by $4\times 4$ matrices with real entries for the real subcases,  $8\times 8$ matrices with real entries for the complex subcases and $16\times 16$ matrices with real entries for the quaternionic subcases. \par
For each inequivalent case the multiplication table of the superdivision algebras generators can be directly read from a given matrix representation.
\par
{{Concerning the mathematical results of the paper we point out that, despite generalizations of the Brauer group having been investigated in the mathematical literature (see e.g. \cite{cgo,lon,frwa}) the explicit construction
of \zzg superdivision algebras and their relations with \\(split-)complex and (split-)quaternions was not carried out. Furthermore, the specialization to the \zzg case allows to apply these superdivision algebras to Hamiltonians constructed with \zzg parafermions. We illustrate this application in the case of a \zzg parafermionic Hamiltonian possessing both time-reversal and particle-hole symmetries. More comments on this construction are given in the Conclusions.}}
\par
The scheme of the paper is the following. The efficient tool of the ``alphabetic presentation of Clifford algebras" to express real representations of Clifford algebras is briefly recalled in Section {\bf 2}. In Section {\bf 3} the notion of graded superdivision algebra is explained. In Section {\bf 4} we apply the alphabetic framework to recover the $10$-fold way classification of division
and ${\mathbb Z}_2$-graded superdivision algebras.  This paves the way for presenting, in Section {\bf 5}, the core of the paper, namely the classification of the associative \zzg superdivision algebras. {{A connection of \zzg superdivision algebras with a \zzg parafermionic Hamiltonian possessing time-reversal and particle-hole symmetries is discussed in Section {\bf 6}.}}  A brief commentary on the obtained results is given in the Conclusions. The alphabetic presentation of ${\mathbb Z}_2$-graded and \zzg superdivision algebras is introduced in Appendix {\bf A}. 

\section{The alphabetic presentation of Clifford algebras}

It was pointed out in \cite{oku} that real, almost complex and quaternionic Clifford algebras are recovered from real matrix representations by taking into account the respective properties concerning the Schur's lemma. This implies that, while almost complex and quaternionic Clifford algebras can be represented by complex matrices, nothing is lost when using only real matrices. The real representations discussed in \cite{oku} are particularly useful for classification purposes, since they allow to consider at once all three cases of real, complex and quaternionic structures.\par

The irreducible representations of Clifford algebras have been classified in \cite{abs}.
It is well known, see e.g. \cite{oku}, that Clifford algebra generators can be expressed as tensor products of the (complex) Pauli matrices and the $2\times 2$ identity.
In order to get the real representations one should drop the imaginary unit $i$. It follows that real representations of Clifford algebras can be obtained by tensoring the $4$ basis elements of the $2\times 2$ real matrices. One can associate a letter to each one of these $4$ matrices. By dropping for convenience the unnecessary symbol ``$\otimes$" of tensor product, which is understood, a matrix representing a Clifford algebra generator can be expressed by a word in a $4$-letter alphabet. This is the ``alphabetic presentation of Clifford algebras" introduced in \cite{tove}. Any irreducible representation of a Clifford algebra is then expressed as a set of equal-length words written in this alphabet, the constructive algorithms presented in \cite{oku} and \cite{crt} recasted \cite{tove} in this language.\par
We are now briefly detailing this construction since it will be applied in the following, at first to recover the $10$-fold way and later to classify the \zzg superdivision algebras. \par
One associates the four letters with a basis of invertible matrices spanning the vector space of  $2\times 2$ real matrices. In this paper we use the following conventions:
{\small{\bea\label{letters}
&I=\left(\begin{array}{cc}1&0\\0&1\end{array}\right),\qquad X=\left(\begin{array}{cc}1&0\\0&-1\end{array}\right),\qquad
Y=\left(\begin{array}{cc}0&1\\1&0\end{array}\right),\qquad A=\left(\begin{array}{cc}0&1\\-1&0\end{array}\right).\qquad&
\eea
}}
The letter $I$ is chosen because it stands for ``the identity matrix", while $A$ stands for ``the antisymmetric matrix". An $n$-character word written in this $4$-letter alphabet represents a $2^n\times 2^n$ real matrix.  {{The connection with the hermitian Pauli matrices mentioned before is given by $\sigma_1= Y, ~\sigma_2= -i A, ~\sigma_3= X$.}}  \par
There are some useful tips for detecting the matrix structure. For instance:
\\
{\it i})  a word whose initial letter is $I$ or $X$  represents a block-diagonal matrix, while if it starts with $Y$ or $A$ the matrix is block-antidiagonal;\\
{\it ii}) a word containing an even (odd) number of letters $A$ represents a symmetric (antisymmetric) matrix;\\
{\it iii}) two different $X,Y,A$ letters represent mutually anticommuting matrices, making easier to check whether two matrices defined by equal-length
words either commute or anticommute.\par
For illustrative purposes and later convenience we discuss the alphabetic presentation of the quaternions. 
A faithful representation of the three imaginary quaternions $q_i$ ($i=1,2,3$) satisfying
\bea\label{quatmult}
q_iq_j &=&-\delta_{ij} +\varepsilon_{ijk}q_k
\eea
(where $\varepsilon_{ijk}$ is the totally antisymmetric tensor normalized as $\varepsilon_{123}=1$) 
is given by either
\bea\label{quat1}
&{\overline q}_1= IA,\qquad {\overline q}_2=AY,\qquad {\overline q}_3= AX,&
\eea
or by
\bea\label{quat2}
&{\widetilde q}_1= AI,\qquad {\widetilde q}_2=YA,\qquad {\widetilde q}_3= XA.&
\eea
Following the rules mentioned above the $4\times 4$ real matrices expressed by (\ref{quat1}) are given by
{\footnotesize{\bea
&{\overline q}_1=\left(\begin{array}{cccc}0&1&0&0\\-1&0&0&0\\0&0&0&1\\0&0&-1&0\end{array}\right),\qquad {\overline q}_2=\left(\begin{array}{cccc}0&0&0&1\\0&0&1&0\\0&-1&0&0\\-1&0&0&0\end{array}\right),\qquad {\overline q}_3=\left(\begin{array}{cccc}0&0&1&0\\0&0&0&-1\\-1&0&0&0\\0&1&0&0\end{array}\right).&
\eea }}
The three imaginary quaternions ${\overline q}_i$ defined by (\ref{quat1}) satisfy the
\bea
\{{\overline q}_i,{\overline q}_j\}&=& - 2\delta_{ij}\cdot {\mathbb I}_4 \qquad\quad ({\mathbb I}_4:= II)
\eea
relations, making them the gamma matrices (see below) of the $Cl(0,3)$ Clifford algebra. Their quaternionic structure is encoded, see \cite{oku}, in the Schur's lemma, which states that the most general matrix $S$ commuting with the ${\overline q}_i$'s is of quaternionic form, being expressed by
\bea
S= \lambda_0\cdot II+\lambda_1\cdot AI +\lambda_2\cdot YA+\lambda_3\cdot XA \quad &\Rightarrow& ~\quad [S,{\overline q}_i]=0,
\eea
for arbitrary real values $\lambda_J\in {\mathbb R}$, with $J=0,1,2,3$. \par
$S$ is defined in terms of the ${\widetilde q}_j$ matrices entering (\ref{quat2}).
Obviously, the identification of (\ref{quat1}) as defining the imaginary quaternions and of (\ref{quat2}) as defining the associated quaternionic structure can be switched.\par
The multiplication table of the $4$ letters, obtained from the (\ref{letters}) identifications,  is

\bea\label{multi}
&\begin{array}{|c||c|c|c|c|}\hline~_r\backslash^{c}&I&X&Y&A \\\hline \hline I&I&X&Y&A \\ \hline X&X&I&A&Y\\ \hline Y&Y&-A&I&-X \\ \hline  A&A&-Y&X&-I\\ \hline
\end{array}&
\eea
{\bf Table 1:} {\it letters multiplications}. The entries of the table denote the result of the multiplication of the row letters ``$r$" (acting on the left) with the column letters ``$c$".\par
~\par
Our nomenclature of Clifford algebras follows \cite{oku,crt}.  The Clifford algebra $Cl(p,q)$ is the Enveloping algebra, over the ${\mathbb R}$ field, of a set of $n\times n$ gamma matrices $\gamma_I$ $(I=1,2,\ldots , p+q)$ satisfying, for any $I,J$ pair:
\bea\label{gammagamma}
\gamma_I\gamma_J+\gamma_J\gamma_I &=& 2\eta_{IJ} \cdot{\mathbb I}_n.
\eea 
${\mathbb I}_n$ denotes the $n\times n $ identity matrix, while $\eta_{IJ}$ is a pseudo-Euclidean diagonal metric
with signature $+1$ for $I,J=1,\ldots, p$ and $-1$ for $I,J=p+1,\ldots p+q$.\par
The irreducible representation of $Cl(p,q)$ is recovered when $n$ is the minimal integer which allows solutions of (\ref{gammagamma}) (sometimes it is useful, as in the applications to superdivision algebras, to skip the irreducibility requirement).\par
The three $2\times 2$ matrices associated with the letters $X,Y,A$ are the gamma matrices of the $Cl(2,1)$ Clifford algebra; the whole set of four letters (\ref{letters}) is a two-dimensional faithful representation of ${\widetilde{\mathbb H}}$, the algebra of the split-quaternions. 
{{We recall (see \cite{mcc}  and \cite{kuto0}) that  both division and split-division algebras are obtained via the Cayley-Dickson doubling construction, their difference being encoded in a sign.
Given an algebra ${\mathbb A}$ over ${\mathbb R}$ endowed with a 
 multiplication $\cdot$, a conjugation $\ast$ and a quadratic form $N$, the Cayley-Dickson doubled algebra
(denoted as ${\mathbb A}^2$) over ${\mathbb R}$ is defined in terms of the operations in ${\mathbb A}$. They are respectively given by:
\\
{\it i}) ~~ multiplication in ${\mathbb A}^2$:
$(x,y)\cdot (z,w) = (x z+\varepsilon w^\ast y, wx +yz^\ast)$,\\
{\it ii}) ~ conjugation in ${\mathbb A}^2$: $(x,y)^\ast = (x^\ast, -y)$,\\
{\it iii})~ quadratic form in ${\mathbb A}^2$:
$N(x,y) = N(x)-\varepsilon N(y)$,\\
where $\varepsilon =\pm 1$ is a sign. The doubling of the real numbers respectively gives the complex (split-complex) numbers for $\varepsilon =-1$ ($\varepsilon =+1$); the doubling of the complex numbers gives the quaternions (split-quaternions) for  $\varepsilon =-1$ ($\varepsilon =+1$). 
}}

\section{Graded superdivision algebras}

In this Section we recall the notion of graded superdivision algebra.
In a ${\mathbb Z}_2$-graded superdivision algebra the generators are split into even (belonging to the $0$-graded sector) and odd (belonging to the $1$-graded sector). In a \zzg  superdivision algebra the generators are accommodated into $2$ bits of information ($00, 10, 01, 11$ graded sectors). A graded superdivision algebra can
be denoted as ${\mathbb D}^{[p]}$, where the non-negative integer $p$ indicates the ${\mathbb Z}_2^p$-grading. 
In this work we limit to consider the $p=0,1,2$ values. The three ordinary associative division algebras over the reals (${\mathbb R}, {\mathbb C}, {\mathbb H}$) are obtained from $p=0$. For later purposes we can set
\bea\label{d0conventions}
&{\mathbb R} := D^{[0]}_{{\mathbb R};1},\qquad {\mathbb C} := D^{[0]}_{{\mathbb R};2},\qquad {\mathbb H} := D^{[0]}_{{\mathbb R};3}. &
\eea
For $p=1,2$ we respectively have
\bea
{\mathbb D}^{[1]} &=& {\mathbb D}^{[1]}_0\oplus {\mathbb D}^{[1]}_1,\nonumber\\
{\mathbb D}^{[2]} &=& {\mathbb D}^{[2]}_{00}\oplus {\mathbb D}^{[2]}_{01 }\oplus {\mathbb D}^{[2]}_{10}\oplus {\mathbb D}^{[2]}_{11}.
\eea   
A homogeneous element $g$ belongs to a given graded sector ($g\in{\mathbb D}^{[1]}_i$ or $g\in{\mathbb D}^{[2]}_{ij}$,
where $i,j$ take values $0,1$). A multiplication, which respects the grading, is defined. 
Let $g_A, g_B\in {\mathbb D}^{[p]}$ be two homogeneous elements, whose respective gradings are
$i_A,i_B$ for ${\mathbb Z}_2$ and the pairs $(i_A,j_A)$, $(i_B,j_B)$ for ${\mathbb Z}_2\times {\mathbb Z}_2$.
The multiplied element $g_A\cdot g_B\in {\mathbb D}^{[p]}$ is homogeneous and its graded sector,
either $i_{A+B}$ or $(i_{A+B}, j_{A+B})$, is obtained from $mod~ 2$ arithmetics:
\bea\label{mod2sum}
{\mathbb Z}_2:\quad i_{A+B} = i_A+i_B; &\quad&  {\mathbb Z}_2\times{\mathbb Z_2}: \quad (i_{A+B}, j_{A+B}) = (i_A+i_B, j_A+j_B).
\eea 
The unit element $1$ will also be denoted as ``$e_0$" (it belongs to the $0$ sector or, respectively, the $00$ sector). In this paper we assume the multiplication to be associative.\par
In a graded superdivision algebra each nonvanishing homogeneous element  {{(belonging to any $i$ or $ij$ sector)}} is invertible. As a consequence, each graded sector is isomorphic, as vector space, to one of the three vector spaces of real numbers, complex numbers or quaternions. It easily follows that the common dimension (in real counting) of each graded sector of a given superdivision algebra is $1$, $2$ or $4$, depending on the case.\par
Each graded sector is spanned by the respective set of basis vectors $e_J$, $f_J$, $g_J$, $h_J$, that can be assigned according to
\bea\label{sdivconventions}
{\textrm{for}}\quad {\mathbb Z}_2\qquad~&:& \quad e_J\in {\mathbb D}^{[1]}_0,\quad f_J\in {\mathbb D}^{[1]}_1;
\nonumber\\
{\textrm{for}}\quad {\mathbb Z}_2\otimes{\mathbb Z}_2&:&\quad e_J\in {\mathbb D}^{[2]}_{00},\quad f_J\in {\mathbb D}^{[2]}_{01},\quad g_J\in {\mathbb D}^{[2]}_{10},\quad h_J\in {\mathbb D}^{[2]}_{11}.
\eea
Depending on the spanning vector space, the suffix $J$ is restricted to be
\bea\label{sdivspan}
&{\mathbb R}: \quad J=0;\qquad {\mathbb C}:\quad J=0,1; \qquad {\mathbb H}: \quad J=0,1,2,3.&
\eea
A generic homogeneous element $e$ belonging to either the $0$ or the $00$ graded sector is expressed by the
linear combination $e=\sum_J \lambda^J e_J$, where the $\lambda^J$'s ($\lambda^J\in {\mathbb R}$) are real
parameters. This formula is extended in obvious way to homogeneous elements belonging to the other graded sectors. 
As mentioned before, in our conventions $e_0$ will denote the unit element.\par
~\par
{\it Remark}: the following remark will be used later on. In a \zzg superdivision algebra the three graded sectors $01,10,11$ are on equal footing and can be interchanged by permutation. Indeed, by setting
$\alpha=01$, $\beta= 10$, $\gamma=11$, the table of the $mod~2$ sums given in (\ref{mod2sum}) of the ${\mathbb Z}_2\times{\mathbb Z}_2$ grading reads as follows:
\bea\label{alphabetagamma}
&\begin{array}{|c||c|c|c|c|}\hline &00&\alpha&\beta&\gamma\\\hline \hline 00&00&\alpha&\beta&\gamma\\  \hline \alpha&\alpha&00&\gamma&\beta\\ \hline
\beta&\beta&\gamma&00&\alpha\\  \hline \gamma&\gamma&\beta&\alpha&00\\ \hline
\end{array}&
\eea
{\bf Table 2:} {\it mod ${2}$ sums of the ${\mathbf {\mathbb Z}_2}\times {\mathbb Z}_2$ grading.}
The table shows the invariance under permutations of the $\alpha,\beta,\gamma$ sectors.\par
~\par
Each graded superdivision algebra over ${\mathbb R}$ can be presented by normalizing any given generator $g\in {\mathbb D}^{[p]}$ so that its square is, up to a sign, the identity ($g^2=\pm e_0$). This leaves a sign ambiguity ($\pm g$) in the normalization of each generator $g\neq e_0$. The graded superdivision algebras are divided into classes of equivalence based on the following set of transformations:\\
{\it i}) the sign flippings $g\mapsto \pm g$ for the generators $g\neq e_0$,\\
{\it ii}) the permutation of the generators inside each graded sector and\\
{\it iii}) for the ${\mathbb Z}_2\times{\mathbb Z}_2$ grading ($p=2$),  the permutation of the equal footing sectors $10,01,11$.\par
This set of transformations defines the $7$ (and respectively $13$) inequivalent classes of ${\mathbb Z}_2$-graded (${\mathbb Z}_2\times{\mathbb Z}_2$-graded) superdivision algebras.\par
In each class of equivalence, a given multiplication table produced among generators induces equivalent multiplication tables obtained by applying the above three transformations. In the case of the quaternions, for instance, this is tantamount to flip the sign ($\epsilon_{ijk} \mapsto -\epsilon_{ijk}$) of the totally antisymmetric tensor entering (\ref{quatmult}).\par   
A multiplication table is straightforwardly recovered from a faithful matrix representation of the graded superdivision algebra.

\section{The  10-fold way revisited}
Each ${{\mathbb{Z}}_2}$-graded superdivision algebra admits an alphabetic presentation; 
the generators (see Appendix {\bf A}) are expressed by equal-length words in a $4$-letter alphabet and the (\ref{letters}) identification of letters and $2\times 2$ matrices holds. \par
In this Section we present the alphabetic derivation of the $10$-fold way for division and ${\mathbb{Z}}_2$-graded superdivision algebras.  We start at first with the three ordinary division algebras.\par
The field of the real numbers ${\mathbb R}$ is the only exception with the (\ref{letters}) identification of letters with $2\times 2$ matrices, since the unit element $e_0$ is the number $1$; the real numbers can, nevertheless, be accommodated into the scheme by setting $e_0=I$, the two-dimensional identity matrix. \par
The complex numbers ${\mathbb C}$ are alphabetically expressed by the two single-character  words $I$ and $A$ (respectively, the identity element and the imaginary unit). \par
 The quaternions ${\mathbb H}$ can be expressed by four two-character words:
either $II, IA, AY, AX$, see (\ref{quat1}) or $II, AI, YA, XA$, see (\ref{quat2}). The $4\times 4$ matrices corresponding to these two choices are related by a similarity transformation.\par
In terms of the (\ref{d0conventions}) positions we can express the sets of generators as
\bea
I~~\quad\quad ~~&\in&{\mathbb D}^{[0]}_{{\mathbb R};1},\nonumber\\
I, ~A\quad~~~~~ &\in &{\mathbb D}^{[0]}_{{\mathbb R};2},\nonumber\\
II, ~IA,~AX,AY&\in&{\mathbb D}^{[0]}_{{\mathbb R};3}.
\eea
The seven ${\mathbb Z}_2$-graded superdivision algebras ($7=2+3+2$) will be denoted as \par
${\mathbb D}^{[1]}_{{\mathbb R};\ast}~$ for the real series (with $\ast=1,2$),\par
 ${\mathbb D}^{[1]}_{{\mathbb C};\ast}~$ for the complex series (with $\ast=1,2,3$) and \par ${\mathbb D}^{[1]}_{{\mathbb H};\ast}~$ for the quaternionic series (with $\ast=1,2$).

\subsection{The real  {{${\mathbb Z}_2$}}-graded superdivision algebras}

Two inequivalent ${\mathbb Z}_2$-graded superdivision algebras are obtained from the real series. On the basis of 
formulas (\ref{mat22}, \ref{z2sectors}, {\ref{z2even}) in Appendix ${\bf A}$ they can be expressed as
\bea\label{z2real}
 {\mathbb D}^{[1]}_{{\mathbb R};1} &:& \qquad I \in {\mathbb D}^{[1]}_0, \qquad A\in {\mathbb D}^{[1]}_1,\nonumber\\
 {\mathbb D}^{[1]}_{{\mathbb R};2} &:& \qquad I \in {\mathbb D}^{[1]}_0, \qquad Y \in {\mathbb D}^{[1]}_1,
\eea
where ${\mathbb D}^{[1]}_0$ {{and}} ${\mathbb D}^{[1]}_1$ denote the respective even and odd sectors.\par
~\par
Since $A^2=-I$, $Y^2=I$ we have that:\\
~\\
$~$ {\it i}) ${\mathbb D}^{[1]}_{{\mathbb R};1}~$ corresponds to the complex numbers ${\mathbb C}$ endowed with a ${\mathbb Z}_2$-grading;\\
{\it ii}) ${\mathbb D}^{[1]}_{{\mathbb R};2}~$ corresponds to the  split-complex numbers ${\widetilde{{\mathbb C}}}$ endowed with a ${\mathbb Z}_2$-grading.\par

\subsection{The complex {{${\mathbb Z}_2$}}-graded superdivision algebras}

Three inequivalent ${\mathbb Z}_2$-graded superdivision algebras are obtained from the complex series. On the basis of 
formulas (\ref{mat22}, \ref{z2sectors}, \ref{z2even}) in Appendix ${\bf A}$ they can be expressed as
\bea\label{z2complex}
 {\mathbb D}^{[1]}_{{\mathbb C};1} &:& \qquad II,~IA \in {\mathbb D}^{[1]}_0, \qquad AX,~AY\in {\mathbb D}^{[1]}_1,\nonumber\\ {\mathbb D}^{[1]}_{{\mathbb C};2} &:& \qquad I I,~IA\in {\mathbb D}^{[1]}_0, \qquad YX,~YY \in {\mathbb D}^{[1]}_1,\nonumber\\
 {\mathbb D}^{[1]}_{{\mathbb C};3} &:& \qquad II,~IA \in {\mathbb D}^{[1]}_0, \qquad ~ AI, ~AA \in {\mathbb D}^{[1]}_1,
\eea
where ${\mathbb D}^{[1]}_0$ (${\mathbb D}^{[1]}_1$) denote the respective even and odd sectors.\par
The alternative presentation $ II,~IA \in {\mathbb D}^{[1]}_0,$ $ YI, ~YA \in {\mathbb D}^{[1]}_1$ produces a superdivision algebra which is isomorphic to $ {\mathbb D}^{[1]}_{{\mathbb C};3} $.\par

We have that:\\

~
\\
$~~$ {\it i}) ${\mathbb D}^{[1]}_{{\mathbb C};1}~$ corresponds to a ${\mathbb Z}_2$-grading of the quaternions ${\mathbb H}$, realizing a graded representation of the $Cl(0,3)$ Clifford algebra;\\
$~$ {\it ii}) ${\mathbb D}^{[1]}_{{\mathbb C};2}~$ corresponds to a ${\mathbb Z}_2$-grading of the split-quaternions ${\widetilde{\mathbb H}}$, realizing a $4\times4$  graded matrix representation of the $Cl(2,1)$ Clifford algebra; \\
{\it iii}) ${\mathbb D}^{[1]}_{{\mathbb C};3}~$ corresponds to a ${\mathbb Z}_2$-grading of an algebra of commuting matrices.

\subsection{The quaternionic {{${\mathbb Z}_2$}}-graded superdivision algebras}
Two inequivalent ${\mathbb Z}_2$-graded superdivision algebras are obtained from the quaternionic series. On the basis of 
formulas (\ref{mat22}, \ref{z2sectors}, \ref{z2even}) in Appendix ${\bf A}$ they can be expressed as
\bea\label{z2quat}
 {\mathbb D}^{[1]}_{{\mathbb H};1} &:& \qquad  III, IIA, IAY, IAX\in {\mathbb D}^{[1]}_0, \qquad AII, AIA, AAY, AAX\in {\mathbb D}^{[1]}_1,\nonumber\\ {\mathbb D}^{[1]}_{{\mathbb H};2} &:& \qquad 
 III, IIA, IAY, IAX \in {\mathbb D}^{[1]}_0, \qquad YII, YIA, YAY, YAX \in {\mathbb D}^{[1]}_1,
\eea
where ${\mathbb D}^{[1]}_0$ (${\mathbb D}^{[1]}_1$) denote the respective even and odd sectors.\par
~\par
The inequivalence of the two superdivision algebras given above  is spotted by taking the squares of the odd generators.
They produce, up to a sign, the $8\times 8$ identity matrix. The signs are $(-+++)$  for 
$ {\mathbb D}^{[1]}_{{\mathbb H};1}$ and  $(+---)$  for 
$ {\mathbb D}^{[1]}_{{\mathbb H};2}$ .\par
~\par
In this Section we recovered, within the alphabetic presentation, the $7$ superdivision algebras discussed in \cite{bae} and presented, under a different name, in \cite{frmo}.

\section{The  13 inequivalent  {{${{{\mathbb Z}_2\times {\mathbb Z}_2}}$}}-graded superdivision algebras}
The alphabetic construction of the \zzg superdivision algebras has been discussed in Appendix {\bf A}. We present here the results. We obtain $13$ inequivalent cases (the defining classes of equivalence have been introduced in Section {\bf 3}). A representative, for each class of equivalence, is given below.\par
The $13$ cases are split into $13=4+5+4$ subcases; $4$ subcases are obtained from the real series, $5$ subcases from the complex series, $4$ subcases from the quaternionic series. The $13$ \zzg superdivision algebras will be named as follows:\par
~\par
${\mathbb D}^{[2]}_{{\mathbb R};\ast}~$ for the real series (with $\ast=1,2,3,4$),\par
 ${\mathbb D}^{[2]}_{{\mathbb C};\ast}~$ for the complex series (with $\ast=1,2,3,4,5$) and \par ${\mathbb D}^{[2]}_{{\mathbb H};\ast}~$ for the quaternionic series (with $\ast=1,2,3,4$).\par
~\par
As mentioned in Appendix {\bf A}, see formula (\ref{ssubalg}), any \zzg superdivision algebra can be characterized by its subalgebras ${\mathbb S}_{01}, {\mathbb S}_{10}, {\mathbb S}_{11}$, obtained by restricting the generators to, respectively, the sectors $00 \& 01$,  $00\& 10$, $00\& 11$. The subalgebras $ {\mathbb S}_{01}, {\mathbb S}_{10}, {\mathbb S}_{11}$ are ${\mathbb Z}_2$-graded superdivision algebras.  Since, as recalled in Section {\bf 3}, see table ({\ref{alphabetagamma}}), the sectors $01,10,11$ are on equal footing, the ${\mathbb Z}_2$-graded subalgebra projections of a \zzg superdivision algebra can be characterized by the triple
\bea\label{sabc}
&({\mathbb S}_\alpha/{\mathbb S}_\beta/{\mathbb S}_\gamma),&
\eea
where the order of the subalgebras is inessential.\par
In the following subsections we separately present the \zzg superdivision algebras for, respectively, the real, complex and quaternionic series.

\subsection{The real {{${\mathbb Z}_2\times{\mathbb Z}_2$-graded}} superdivision algebras}
The four inequivalent \zzg superdivision algebras ${\mathbb D}^{[2]}_{{\mathbb R};\ast}~$ of the real series possess four generators. The matrix representatives of each class of equivalence can be expressed as in the table below, which gives the matrix generator of the corresponding graded sector:
~\par
\bea\label{fourz2z2r}
&\begin{array}{|c|c|c|c|c|}\hline &00&01&10&11\\ \hline {\mathbb D}^{[2]}_{{\mathbb R};1}:&II&IA&AX&AY\\  \hline
{\mathbb D}^{[2]}_{{\mathbb R};2}:&II&IA&YX&YY\\  \hline
{\mathbb D}^{[2]}_{{\mathbb R};3}:&II&IA&AI&AA\\  \hline{\mathbb D}^{[2]}_{{\mathbb R};4}:&II&IY&YI&YY\\  \hline
\end{array}&
\eea
{\bf Table 3:} {\it presentation of the ${\mathbf 4}$ inequivalent ${\mathbf{\mathbb Z}_2\times{\mathbb Z}_2}$-graded superdivision algebras of the real series.}\\
~\par
Some comments are in order.\\
~\\
{\it Comment} $1$ - squaring the matrices entering the $01,10,11$ sectors gives the signs
{{\bea
& {\mathbb D}^{[2]}_{{\mathbb R};1}:~---; \quad ~~{\mathbb D}^{[2]}_{{\mathbb R};2}:~-++;\quad~~ {\mathbb D}^{[2]}_{{\mathbb R};3}:~--+;\quad~~
 {\mathbb D}^{[2]}_{{\mathbb R};4}:~+++.&
\eea}}
{\it Comment} $2$ - 
the projections to the ${\mathbb Z}_2$-graded subalgebras, see formula (\ref{sabc}), are given by
\bea
&{\mathbb D}^{[2]}_{{\mathbb R};1}: ({\underline 1}/{\underline 1}/{\underline 1});  \qquad
{\mathbb D}^{[2]}_{{\mathbb R};2}: ({\underline 1}/{\underline 2}/{\underline 2});  \qquad
{\mathbb D}^{[2]}_{{\mathbb R};3}: ({\underline 1}/{\underline 1}/{\underline 2});  \qquad
{\mathbb D}^{[2]}_{{\mathbb R};4}: ({\underline 2}/{\underline 2}/{\underline 2}),&
\eea
where, for simplicity, we denoted ${\underline 1}:=
{\mathbb D}^{[1]}_{{\mathbb R};1}$ and $  {\underline 2}:=
{\mathbb D}^{[1]}_{{\mathbb R};2}.$\\
{\it Comment} $3$ -  the ${\mathbb D}^{[2]}_{{\mathbb R};1}$ superdivision algebra is a ${\mathbb Z}_2\times{\mathbb Z}_2$ gradation of the quaternions ${\mathbb H}$, 
  the ${\mathbb D}^{[2]}_{{\mathbb R};2}$ superdivision algebra is a ${\mathbb Z}_2\times{\mathbb Z}_2$ gradation
 of the split-quaternions ${\widetilde {\mathbb H}}$,  the superdivision algebras ${\mathbb D}^{[2]}_{{\mathbb R};3}$ and ${\mathbb D}^{[2]}_{{\mathbb R};4}$ are commutative.

\subsection{The complex {{${\mathbb Z}_2\times{\mathbb Z}_2$-graded}} superdivision algebras}
The  \zzg superdivision algebras ${\mathbb D}^{[2]}_{{\mathbb C};\ast}~$ of the complex series possess eight generators. The multiplication tables of the admissible alphabetic presentations are grouped into $5$ classes of equivalence. The matrix representatives of each class of equivalence can be expressed as in the table below, which gives the matrix generators of the corresponding graded sector:
~\par
\bea\label{fivez2z2c}
&\begin{array}{|c|c|c|c|c|}\hline &00&01&10&11\\ \hline {\mathbb D}^{[2]}_{{\mathbb C};1}:&III, ~IIA&IAX, ~IAY&AIX,~AIY&AAI, ~AAA\\  \hline
{\mathbb D}^{[2]}_{{\mathbb C};2}:&III,~IIA&AIX,~AIY&IYX,~IYY&AYI, ~AYA\\  \hline
{\mathbb D}^{[2]}_{{\mathbb C};3}:&III,~IIA&YIX,~YIY&IYX,~IYY&YYI, ~YYA\\  
\hline{\mathbb D}^{[2]}_{{\mathbb C};4}:&III,~IIA&YII,~YIA&XYI,~XYA&AYI,~AYA\\  \hline
{\mathbb D}^{[2]}_{{\mathbb C};5}:&III,~IIA&YII,~YIA&IYI,~IYA&YYI,~YYA\\  \hline
\end{array}&
\eea
{\bf Table 4:} {\it presentation of the ${\mathbf 5}$ inequivalent ${\mathbf{\mathbb Z}_2\times{\mathbb Z}_2}$-graded superdivision algebras of the complex series.}\\
~\\
Following the construction outlined in Appendix {\bf A}, by setting the ${\mathbb Z}_2$-graded superdivision algebras
${\mathbb S}_{01}$ (obtained from the $00$ and $01$ sectors) and ${\mathbb S}_{10}$ (obtained from the $00$ and $10$ sectors), the 
${\mathbb Z}_2$-graded superdivision algebra ${\mathbb S}_{11}$ is determined.  It is convenient to denote here
the three complex ${\mathbb Z}_2$-graded superalgebras as ${\underline 1}:=
{\mathbb D}^{[1]}_{{\mathbb C};1},~$  $  {\underline 2}:=
{\mathbb D}^{[1]}_{{\mathbb C};2}~$ and $~  {\underline 3}:=
{\mathbb D}^{[1]}_{{\mathbb C};3}$.\\
{{The projections to the ${\mathbb Z}_2$-graded subalgebras are easily obtained. From (\ref{fivez2z2c}) one gets, e.g., that the $ {\mathbb D}^{[2]}_{{\mathbb C};1}$ subalgebras are:\\ ${\mathbb S}_{01}={\underline 1}$ 
(obtained from the $III, IIA$ and $IAX, IAY$ nonminimal representation of  ${\mathbb D}^{[1]}_{{\mathbb C};1}={\underline 1}$), \\${\mathbb S}_{10}={\underline 1}$ 
(obtained from the $III, IIA$ and $AIX, AIY$ nonminimal representation of  ${\mathbb D}^{[1]}_{{\mathbb C};1}={\underline 1}$), \\
${\mathbb S}_{11}={\underline 3}$ 
(obtained from the $III, IIA$ and $AAI, AAA$ nonminimal representation of  ${\mathbb D}^{[1]}_{{\mathbb C};3}={\underline 3}$).}}\par
By taking into account, see formula (\ref{sabc}), that the $01,10,11$ sectors are on equal footing, the projections to ${\mathbb Z}_2$-graded superdivision algebras are given by
\bea
&
{\mathbb D}^{[2]}_{{\mathbb C};1}: ({\underline 1}/{\underline 1}/{\underline 3});\qquad
{\mathbb D}^{[2]}_{{\mathbb C};2}: ({\underline 1}/{\underline 2}/{\underline 3});  \qquad
{\mathbb D}^{[2]}_{{\mathbb C};3}: ({\underline 2}/{\underline 2}/{\underline 3});  \qquad
{\mathbb D}^{[2]}_{{\mathbb C};4}: ({\underline 3}/{\underline 3}/{\underline 3});\qquad
{\mathbb D}^{[2]}_{{\mathbb C};5}: ({\underline 3}/{\underline 3}/{\underline 3}).&\nonumber\\&&
\eea
~\\
{\it Remark}:
the inequivalent \zzg superdivision algebras ${\mathbb D}^{[2]}_{{\mathbb C};4},~ {\mathbb D}^{[2]}_{{\mathbb C};5}$ are not discriminated by their ${\mathbb Z}_2$-graded projections. Their difference is spotted as follows:\\
{\it i}) in the ${\mathbb D}^{[2]}_{{\mathbb C};4}$ superdivision algebra any pair of $g, g'$ generators belonging to different $01,10,11$ graded sectors {\it anticommute};\\
{\it ii}) in the ${\mathbb D}^{[2]}_{{\mathbb C};5}$ superdivision algebra any pair of $g, g'$ generators belonging to different $01,10,11$ graded sectors {\it commute}.

\subsection{The quaternionic {{${\mathbb Z}_2\times{\mathbb Z}_2$-graded}} superdivision algebras}
The four inequivalent \zzg superdivision algebras ${\mathbb D}^{[2]}_{{\mathbb H};\ast}~$ of the quaternionic series possess sixteen generators. The matrix representatives of each class of equivalence are given below. In all four cases the generators of the $00$ sector can be given by
\bea\label{z2z2quata}
00&:& ~ IIII,~IIIA,~IIAX,~IIAY.
\eea
The generators of the $01$, $10$, $11$  sectors are expressed as
~\par
{\small{\bea\label{z2z2quatb}
&\begin{array}{|c|c|c|c|}\hline &01&10&11\\ \hline {\mathbb D}^{[2]}_{{\mathbb H};1}:& IAII,~IAIA,~IAAX,~IAAY&AXII,~AXIA,~AXAX,~AXAY&AYII,~AYIA,~AYAX,~AYAY\\  \hline
{\mathbb D}^{[2]}_{{\mathbb H};2}:&IAII,~IAIA,~IAAX,~IAAY&AIII,~AIIA,~AIAX,~AIAY&AAII~,AAIA~,AAAX,~AAAY\\  \hline
{\mathbb D}^{[2]}_{{\mathbb H};3}:&IAII,~IAIA,~IAAX,~IAAY&YXII,~YXIA,~YXAX,~YXAY&YYII~,~YYIA,~YYAX,~YYAY\\  \hline{\mathbb D}^{[2]}_{{\mathbb H};4}:&IYII,~IYIA,~IYAX,~IYAY&YIII,~YIIA,~YIAX,~YIAY&YYII,~YYIA,~YYAX,~YYAY\\  \hline
\end{array}&\nonumber\\&&
\eea}}
{\bf Table 5:} {\it presentation of the ${\mathbf 4}$ inequivalent ${\mathbf{\mathbb Z}_2\times{\mathbb Z}_2}$-graded superdivision algebras of the quaternionic series.}\\
~\\
Let us redefine ${\underline 1}$, ${\underline 2}$ as the quaternionic ${\mathbb Z}_2$-graded superdivision algebras, so that,  ${\underline 1} :=  {\mathbb D}^{[1]}_{{\mathbb H};1}$ and ${\underline 2} :=  {\mathbb D}^{[1]}_{{\mathbb H};2}$. The quaternionic projections, see formula (\ref{sabc}), are then given by
\bea
&
{\mathbb D}^{[2]}_{{\mathbb H};1}: ({\underline 1}/{\underline 1}/{\underline 1});\qquad
{\mathbb D}^{[2]}_{{\mathbb H};2}: ({\underline 1}/{\underline 1}/{\underline 2});  \qquad
{\mathbb D}^{[2]}_{{\mathbb H};3}: ({\underline 1}/{\underline 2}/{\underline 2});  \qquad
{\mathbb D}^{[2]}_{{\mathbb H};4}: ({\underline 2}/{\underline 2}/{\underline 2}).
\eea
~\par
{\it Comment}:  the $13$ inequivalent multiplication tables of the \zzg superdivision algebras are straightforwardly recovered from the matrix representations of the generators, given in  (\ref{fourz2z2r}) for the real series, (\ref{fivez2z2c}) 
for the complex series and ({\ref{z2z2quata},\ref{z2z2quatb}) for the quaternionic series. To save space they will not be reported here.\par

\section{{{{\bf{\Large{An application to a ${\mathbb Z}_2\times{\mathbb Z}_2$-graded parafermionic Hamiltonian}}}}}}
We illustrate here  how ${\mathbb Z}_2\times{\mathbb Z}_2$-graded superdivision algebras can be applied in investigating the properties of ${\mathbb Z}_2\times{\mathbb Z}_2$-graded parafermionic Hamiltonians. In the example given below we consider a single-particle type of Hamiltonian; following the scheme of \cite{{top1},{top2}}, noninteracting multi-particle Hamiltonians (which respect the ${\mathbb Z}_2\times{\mathbb Z}_2$-graded parastatistics) are constructed  via first quantization of single-particle Hamiltonians.\par
In application to condensed matter physics and the relation with the periodic table two broad classes of Hamiltonians are usually investigated, see  \cite{{rsfl},{mofu},{zir2}}: the fermionic-type of Hamiltonians presented in \cite{alzi} in connection with random matrix ensembles and the Dirac-type Hamiltonians. For our illustrative purposes let's discuss 
here a construction based on a (para)fermionic Hamiltonian.\par
A fermionic oscillator described by $\beta, \gamma=\beta^\dagger$  is introduced through the {{$2\times 2$ matrices}}
{\small{
\bea\label{bc}
&\beta =\left(\begin{array}{cc} 0&1\\0&0\end{array}\right),\quad ~ \gamma =\left(\begin{array}{cc} 0&0\\1&0\end{array}\right),&
\eea
}}
so that
\bea\label{ferosc}
&\beta^2=\gamma^2=0, \qquad \beta\gamma+\gamma\beta= {\mathbb I}_2.&
\eea
By setting
\bea\label{pairoffer}
&\psi_1=\beta\otimes I, \quad {\psi_1}^\dagger = \gamma \otimes I \quad {\textrm{and}} \quad
\psi_2=X\otimes\beta,\quad  {\psi_2}^\dagger = X\otimes\gamma &
\eea
(where $I,X$ are two letters of the (\ref{letters}) alphabet) we obtain a system of two fermions satisfying the superalgebra
\bea
&\psi_1^2=\psi_2^2=(\psi_1^\dagger)^2=(\psi_2^\dagger)^2=0, \quad \{\psi_1,\psi_1^\dagger\}=\{\psi_2,\psi_2^\dagger\}={\mathbb I}_4,&\nonumber\\
&\{\psi_1,\psi_2\}=\{\psi_1,\psi_2^\dagger\}=\{\psi_1^\dagger,\psi_2\}=\{\psi_1^\dagger,\psi_2^\dagger\}=0.&\eea
By setting
{{\bea\label{pairofz2z2pf}
&{\overline\psi}_1=\psi_1=\beta\otimes I, \quad {{\overline \psi}_1}^\dagger = {{\psi}_1}^\dagger=\gamma \otimes I \quad {\textrm{and}} \quad
{\overline\psi}_2=I\otimes\beta,\quad  {{\overline \psi}_2}^\dagger = I\otimes\gamma &
\eea}}
we obtain a system of two ${\mathbb Z}_2\times{\mathbb Z}_2$-graded parafermions which are accommodated, see (\ref{fromz2toz2z2}), in the graded sectors of the $4\times 4$ matrices according to ${\overline\psi}_1,{\overline\psi}_1^\dagger\in M_{10}$ and  ${\overline\psi}_2,{\overline\psi}_2^\dagger\in M_{01}$.   These parafermions satisfy the the ${\mathbb Z}_2\times{\mathbb Z}_2$-graded superalgebra
\bea
&{\overline\psi}_1^2={\overline\psi}_2^2=({\overline\psi}_1^\dagger)^2=({\overline\psi}_2^\dagger)^2=0, \quad 
\{{\overline \psi}_1,{\overline \psi}_1^\dagger\}=\{{\overline\psi}_2,{\overline \psi}_2^\dagger\}={\mathbb I}_4,&\nonumber\\
&\relax [{\overline \psi}_1,{\overline \psi}_2]=[{\overline \psi}_1,{\overline \psi}_2^\dagger]=[{\overline \psi}_1^\dagger,{\overline \psi}_2]=[{\overline \psi}_1^\dagger,{\overline \psi}_2^\dagger]=0.&\eea
{{The most general bosonic and self-adjoint $4\times 4$ Hamiltonian $H$, expressed in terms of the pair of (\ref{pairoffer}) fermionic oscillators, depends on $8$ real parameters; it is given by }}
{\small{\bea\label{fermionicham}
H&=& c_1\cdot {\mathbb I}_4 +c_2\cdot \psi_1^\dagger\psi_1+c_3\cdot\psi_2^\dagger\psi_2+c_4\cdot\psi_1^\dagger\psi_1\psi_2^\dagger\psi_2+c_5\cdot \psi_1^\dagger\psi_2+c_5^\ast\cdot \psi_2^\dagger\psi_1+c_6\cdot\psi_1^\dagger\psi_2^\dagger+c_6^\ast\cdot \psi_2\psi_1,\nonumber\\
\eea}}
where $c_1,c_2,c_3,c_4$ are arbitrary real coefficients. The arbitrary complex parameters $c_5,c_6$ produce off-diagonal terms. \par
By taking into account, see \cite{top1}, that in a ${\mathbb Z}_2\times{\mathbb Z}_2$-graded theory the observables are superselected to belong to the $00$-sector,   the most general $4\times 4$ Hamiltonian ${\overline H}$ expressed in terms of a pair of parafermionic oscillators (\ref{pairofz2z2pf}) is given by the diagonal operator
\bea\label{pfham}
{\overline H}&=& c_1\cdot {\mathbb I}_4 +c_2\cdot {\overline \psi}_1^\dagger{\overline \psi}_1+c_3\cdot {\overline \psi}_2^\dagger{\overline \psi}_2+c_4\cdot{\overline \psi}_1^\dagger{\overline \psi}_1{\overline \psi}_2^\dagger{\overline \psi}_2,
\eea
for $c_1,c_2,c_3,c_4$ arbitrary real coefficients; we have ${\overline H}= diag(c_1,c_1+c_3,c_1+c_2,c_1+c_2+c_3+c_4)$. \par
By setting $c_5=c_6=0$ in (\ref{fermionicham}), the two Hamiltonians coincide (${\overline H}=H$). This is a general feature, already observed in \cite{brdu}, of ${\mathbb Z}_2\times{\mathbb Z}_2$-graded theories. ${\mathbb Z}_2\times{\mathbb Z}_2$-graded parafermionic models are a subclass of fermionic models. In their common range of parameters {\it single-particle} Hamiltonians, as the one entering ({\ref{fermionicham},\ref{pfham}), admit two different but equivalent (the fermionic and  the  ${\mathbb Z}_2\times{\mathbb Z}_2$-graded parafermionic) descriptions. The situation, as shown in \cite{top1,top2},  drastically changes for the {\it multi-particle} Hamiltonians
 induced by first-quantization of systems like ({\ref{fermionicham},\ref{pfham}). The induced (para)statistics  imply inequivalent quantum models  in their multi-particle sectors.\par
We investigate now under which conditions the Hamiltonian (\ref{pfham}) admits both time-reversal and particle-hole symmetry. We use the conventions of \cite{mofu}. The requirement is to find three operators $J,T,C$ which satisfy the following set of properties. $J$ is the imaginary unit which defines the complex structure, while $T$ and $C$ are anti-unitary operators.
$T,C$ respectively define the time-reversal and the particle-hole symmetry. The algebraic conditions to be satisfied
are
\bea\label{quadratic}
&J^2=-{\mathbb I}_4, \quad T^2=\varepsilon_T\cdot{\mathbb I}_4,\quad C^2=\varepsilon_C\cdot{\mathbb I}_4&
\eea
($T,C$ are normalized up to a sign, namely $\varepsilon_T,\varepsilon_C =\pm1$).\par
The compatibility conditions for $J,T,C$ are
\bea\label{compatibility}
&\{T,J\}=\{C,J\}=[T,C]=0.&
\eea 
The symmetry properties of a diagonal Hamiltonian ${\widetilde H}$ require (see \cite{mofu,zir2}) the vanishing of the
(anti)commutators
\bea\label{symmetry}
&[J,{\widetilde H}]= [T,{\widetilde H}]=\{C,{\widetilde H}\}=0.&
\eea
We apply the alphabetic presentation to derive the $J,T,C$ operators satisfying (\ref{quadratic},\ref{compatibility}).
Concerning the Hamiltonian ${\widetilde H}$ we discuss its two different interpretations, either as a bosonic element of a ${\mathbb Z}_2$-graded superalgebra or as a $00$-graded element of a ${\mathbb Z}_2\times{\mathbb Z}_2$-graded superalgebra. Due to the (\ref{pairoffer}) positions, in the ${\mathbb Z}_2$-graded description the non-vanishing entries of the $4\times 4$ real bosonic  matrices are accommodated into
 {\footnotesize{$ \left(\begin{array}{cccc}\ast&0&0&\ast\\ 0&\ast&\ast&0\\0&\ast&\ast&0\\ \ast&0&0&\ast \end{array}\right)
$
}}. In the ${\mathbb Z}_2\times{\mathbb Z}_2$-graded description the antidiagonal elements belong to the $11$-sector.
\par
The imaginary unit $J$ should be an even element; without loss of generality we can assume it to be given by the $2$-letter word
\bea
J&=& YA.
\eea
In the alphabetic presentation the candidates for the $T,C$ operators which anticommute with $J$ are given by the $8$ operators $IX, IY, YX, YY, XI, AI, XA, AA$.\par
The requirement $[J,{\widetilde H}]=0$ for the Hamiltonian implies that ${\widetilde H}= (c_1,c_1+c_2,c_1+c_2,c_1)$. The implementation of the $\{C, {\widetilde H}\}$ particle-hole symmetry further sets the nonvanishing  Hamiltonian to be given, up to a normalizing factor, by
\bea\label{hamilt}
{\widetilde H}&=& XX~=~diag(1,-1,-1,1).
\eea
The alphabetic pairs of $C,T$ operators satisfying (\ref{quadratic},\ref{compatibility},\ref{symmetry}) are given by
\bea
BDI ~{\textrm{class}}:~~\begin{array}{|c|c|}\hline C&T\\ \hline IY&YY\\ \hline IY&XI \\ \hline YX&IX \\ \hline YX&AA \\ \hline  \end{array} 
\qquad \qquad CI ~{\textrm{class}}:~~\begin{array}{|c|c|}\hline C&T\\ \hline XA&YY\\ \hline XA&XI \\ \hline AI&IX \\ \hline AI&AA \\ \hline  \end{array} 
\eea
{\bf Table 6:} {\it admissible pairs of particle-hole and time-reversal operators for the Hamiltonian ${\widetilde H}$.}\\
~\\
In all above cases $T^2={\mathbb I}_4$ so that $\varepsilon_T=+1$;  the $C$ operators entering the left table
are such that $C^2 ={\mathbb I}_4$ with $\varepsilon_C=+1$, while the $C$ operators entering the right table are such that $C^2=-{\mathbb I}_4$ with $\varepsilon_C=-1$.\par
Depending on the choice of the symmetry operators, the (\ref{hamilt}) Hamiltonian falls into both the $BDI$ and the $CI$
classes of symmetry under time-reversal and particle-hole symmetries (a discussion of these classes  can be found, e.g., in\cite{mofu}). They correspond to a Hamiltonian invariant under time-reversal and particle-hole symmetries with, respectively, the normalizations $\varepsilon_C=\varepsilon_T=1$ ($BDI$ class) and $\varepsilon_C=-1,~\varepsilon_T=1$
($CI$ class).\par
Given pairs of the $J,C,T$ operators induce graded superdivision algebras which differ depending on the fermionic or ${\mathbb Z}_2\times {\mathbb Z}_2$-graded parafermionic interpretation of ${\widetilde H}=XX$. Let's analyze the various cases:\\
$~~$ {\it i}) the pair $J, T$ induces the closure, 
under multiplication, of the $4$ operators ${\mathbb I}_4, J, T, JT$;\\
$~$ {\it ii}) the pair $J, C$ induces the closure, under multiplication, of the $4$ operators ${\mathbb I}_4, J, C, JC$;\\
{\it iii}) the pair $T, C$ induces the closure, under multiplication, of the $4$ operators ${\mathbb I}_4, T, C, TC$.\par
The third case is particularly relevant because it involves the time-reversal and particle-hole symmetry operators.\par
The following structures can be observed (the ${\mathbb Z}_2$- and \zzg superdivision algebras presented below are denoted following the notation given in formulas (\ref{z2complex},\ref{fourz2z2r})).\par
~\par
{\bf Case} $i$: for no choice of $T$ the four operators ${\mathbb I}_4, J, T, JT$ have the structure of a graded superdivision algebra; the choice $T=YY$ implies, e.g., that $II={\mathbb I}_4$ and $YY$ belong to the even sector of a ${\mathbb Z}_2$-graded superalgebra. The product $(II+YY)\cdot(II-YY)=0$ produces, on the other hand, divisors of zero. Case {\it i} fails to produce ${\mathbb Z}_2$- and \zzg superdivision algebras.\par
~\par
{\bf Case} $ii$:  two inequivalent sets of ${\mathbb I}_4, J, C, JC$ operators are recovered for the $BDI$ and $CI$ classes of symmetry. We can set,
without loss of generality,
{{\bea
BDI ~{\textrm{class}}: II, YA, IY, YX\quad &{\textrm{and}}&\quad CI~ {\textrm{class}}: II, YA, AI, XA.
\eea}}
  The $BDI$ class corresponds to the set of split-quaternions; the $CI$ class corresponds to the set of quaternions.\\
In the ${\mathbb Z}_2$-graded fermionic interpretation of the ${\widetilde H}$ Hamiltonian we obtain the
respective ${\mathbb Z}_2$-graded superdivision algebras:
\bea
&BDI ~{\textrm{class}}:~~ II, YA \in [0],~~ IY, YX \in [1] \Rightarrow  
{\mathbb D}^{[1]}_{{\mathbb C};2}~{\textrm{superdivision algebra}};
&\nonumber\\
&~~CI ~{\textrm{class}}:~~ II, YA \in [0],~~ AI, XA \in [1] \Rightarrow  
{\mathbb D}^{[1]}_{{\mathbb C};1}~{\textrm{superdivision algebra}}.
&
\eea
In the ${\mathbb Z}_2\times {\mathbb Z}_2$-graded parafermionic interpretation of the ${\widetilde H}$ Hamiltonian the same set of operators are accommodated into the
${\mathbb Z}_2\times{\mathbb Z}_2$-graded superdivision algebras:
\bea
&BDI ~{\textrm{class}}:~~ II\in [00], ~YA \in [11],~ IY\in[01],~ YX \in [10] \Rightarrow  
{\mathbb D}^{[2]}_{{\mathbb R};2}~{\textrm{superdivision algebra}};
&\nonumber\\
&~~CI ~{\textrm{class}}:~~ II\in [00], ~YA \in [11],~ XA\in[01],~ AI \in [10] \Rightarrow  
{\mathbb D}^{[2]}_{{\mathbb R};1}~{\textrm{superdivision algebra}}.
&\nonumber\\
&&
\eea

The results of case {\it ii}  can be summarized in the table
\bea
&\begin{array}{|c|c|c|}\hline  &{\mathbb Z}_2&{\mathbb Z}_2\times{\mathbb Z}_2 \\ \hline BDI:&{\mathbb D}^{[1]}_{{\mathbb C};2} &{\mathbb D}^{[2]}_{{\mathbb R};2} \\ 
\hline CI:&{\mathbb D}^{[1]}_{{\mathbb C};1} &{\mathbb D}^{[2]}_{{\mathbb R};1} \\ \hline  \end{array} &
\eea
{\bf Table 7:} {\it graded superdivision algebras induced by the $J,C$ operators.}\\
~\par
{\bf Case $iii$}:  two inequivalent sets of ${\mathbb I}_4, T, C, TC$ operators are recovered for the $BDI$ and $CI$ classes of symmetry. We can set
\bea
BDI ~{\textrm{class}}: II, YY, IY, YI\quad &{\textrm{and}}&\quad CI~ {\textrm{class}}: II, AA, AI, IA.
\eea
In the ${\mathbb Z}_2$-graded fermionic interpretation of the ${\widetilde H}$ Hamiltonian these operators fail to produce a ${\mathbb Z}_2$-graded superdivision algebra.\par
In the ${\mathbb Z}_2\times {\mathbb Z}_2$-graded parafermionic interpretation of the ${\widetilde H}$ Hamiltonian the same set of operators are accommodated into the
${\mathbb Z}_2\times{\mathbb Z}_2$-graded superdivision algebras:
\bea
&BDI ~{\textrm{class}}:~~ II\in [00], ~YY \in [11],~ IY\in[01],~ YI \in [10] \Rightarrow  
{\mathbb D}^{[2]}_{{\mathbb R};4}~{\textrm{superdivision algebra}};
&\nonumber\\
&~~CI ~{\textrm{class}}:~~ II\in [00], ~AA \in [11],~ IA\in[01],~ AI \in [10] \Rightarrow  
{\mathbb D}^{[2]}_{{\mathbb R};3}~{\textrm{superdivision algebra}}.
&\nonumber\\
&&
\eea
This result can be summarized in the table
\bea
&\begin{array}{|c|c|c|}\hline  &{\mathbb Z}_2&{\mathbb Z}_2\times{\mathbb Z}_2 \\ \hline BDI:&-&{\mathbb D}^{[2]}_{{\mathbb R};4} \\ 
\hline CI:&-&{\mathbb D}^{[2]}_{{\mathbb R};3} \\ \hline  \end{array} &
\eea
{\bf Table 8:} {\it graded superdivision algebras induced by the $T,C$ operators.}\\
~\\
It was already observed in \cite{aktt1,aktt2} that ${\mathbb Z}_2\times{\mathbb Z}_2$-graded superalgebras can accommodate symmetry operators (in that case of Partial Differential Equations) which do not fit into the ${\mathbb Z}_2$-graded superalgebra setting.\par
We conclude with a remark. The ${\widetilde H}$ Hamiltonian (\ref{hamilt}) falls into the class of Hamiltonians discussed in \cite{top2} which admit $9$ inequivalent multiparticle quantizations induced by different gradings. Two such gradings are respectively given by the  fermionic oscillators entering (\ref{fermionicham}) and by the parafermionic oscillators entering (\ref{pfham}).

\section{Conclusions}

In this paper we showed that the $7$ inequivalent ${\mathbb Z}_2$-graded superdivision algebras admit an alphabetic presentation; we extended this framework to individuate $13$ classes of equivalence of the ${\mathbb Z}_2\times{\mathbb Z}_2$-graded superdivision algebras. This result has implications both in mathematics and in physical applications.  On the mathematical side we recall the investigations of \zzg structures (see, e.g., \cite{brgr} for a state of the art account of Riemannian structures on \zzg manifolds). On the physical side, the seemingly most promising application is the extension of the periodic table of topological insulators and superconductors to accommodate a ${\mathbb Z}_2\times {\mathbb Z}_2$ grading. {{It is beyond the scope of this paper to present such investigation which is left for a future work based on the notion of  \zzg supercommutants 
(that is, the algebras of \zzg matrices which (anti)commute with the \zzg superdivision algebras). In this paper we presented an example illustrating how ${\mathbb Z}_2\times{\mathbb Z}_2$-graded superdivision algebras can be applied to parafermionic Hamiltonians with time-reversal and particle-hole symmetries.  We recall that \zzg parafermions are theoretically detectable, see \cite{top1}.  On the experimental side, see the recent work  \cite{exp}, para-oscillators have finally being investigated and engineered in the laboratory.  It seems then feasible, both theoretically and experimentally, to use the alphabetic presentation to engineer the construction of \zzg parafermionic oscillators.  The expectation is that the \zzg parafermionic Hamiltonians induce new features which are not encoded in the $10$-fold way.  }}\par
{{The framework of the alphabetic presentation of Clifford algebras can be extended to determine the classification of ${\mathbb Z}_2^n$-graded superdivision algebras for $n>2$  integral values. The investigation of the $n=3,4,\ldots$ cases is left for future works.}}

~

  \renewcommand{\theequation}{A.\arabic{equation}}
  \setcounter{equation}{0} 
\par
~\\
{\bf{\Large{Appendix A: alphabetic presentation of superdivision algebras}}}
~\par
~\par
We extend here the alphabetic presentation (described in Section {\bf 2} for Clifford algebras) to the cases of ${\mathbb Z}_2$ and \zzg superdivision algebras.\par
Any homogeneous element $g$ of a superdivision algebra is represented by an invertible real matrix which takes the form $g=M\otimes N$, where
the matrix $M$ encodes the information of the grading, either ${\mathbb Z}_2$ or ${\mathbb Z}_2\times{\mathbb Z}_2$, while  the matrix $N$ encodes the information of the real, complex or quaternionic structure. \\
The matrix size for $M$ is
\bea
&{\mathbb Z}_2{\textrm{-grading}}: ~(2\times 2);\qquad \qquad
{\mathbb Z}_2\times
{\mathbb Z}_2{\textrm{-grading}}: ~(4\times 4).&
\eea
The matrix size for $N$ is
\bea
&{\mathbb R}{\textrm{-series}}: ~(1\times 1);\qquad \quad
{\mathbb C}{\textrm{-series}}: ~(2\times 2);\qquad \quad
{\mathbb H}{\textrm{-series}}: ~(4\times 4).&
\eea
Concerning the ${\mathbb Z}_2$ grading, the even (odd) sector is denoted as $M_0$ ($M_1$); the nonvanishing elements are accommodated according to  {\small{\bea\label{mat22}
\left(\begin{array}{cc}\ast&0\\0&\ast\end{array}\right)\in M_0,&\qquad& \left(\begin{array}{cc}0&\ast\\ \ast&0\end{array}\right)\in M_1.
\eea
}}
The tensor products of the ${\mathbb Z}_2$-graded matrices (\ref{mat22}) produce the
 \zzg matrices  $M_{ij}$ ($ij$ denotes the grading) according to

{\footnotesize{\bea\label{fromz2toz2z2}
\left(\begin{array}{cc}\ast&0\\0&\ast\end{array}\right)\otimes \left(\begin{array}{cc}\ast&0\\0&\ast\end{array}\right)\mapsto \left(\begin{array}{cccc}\ast&0&0&0\\ 0&\ast&0&0\\0&0&\ast&0\\ 0&0&0&\ast \end{array}\right)\in M_{00},&&
\left(\begin{array}{cc}\ast&0\\0&\ast\end{array}\right)\otimes \left(\begin{array}{cc}0&\ast\\\ast&0\end{array}\right)\mapsto \left(\begin{array}{cccc}0&\ast&0&0\\ \ast&0&0&0\\0&0&0&\ast\\ 0&0&\ast&0 \end{array}\right)\in M_{01},\nonumber\\
\left(\begin{array}{cc}0&\ast\\ \ast&0\end{array}\right)\otimes \left(\begin{array}{cc}\ast&0\\0&\ast\end{array}\right)\mapsto \left(\begin{array}{cccc}0&0&\ast&0\\ 0&0&0&\ast\\\ast&0&0&0\\ 0&\ast&0&0 \end{array}\right)\in M_{10},&&
\left(\begin{array}{cc}0&\ast\\\ast&0\end{array}\right)\otimes \left(\begin{array}{cc}0&\ast\\\ast&0\end{array}\right)\mapsto \left(\begin{array}{cccc}0&0&0&\ast\\ 0&0&\ast&0\\0&\ast&0&0\\ \ast&0&0&0 \end{array}\right)\in M_{11}.\nonumber\\
&&
\eea
}}
In the ${\mathbb Z}_2$-grading the $M_0$, $M_1$ sectors can be spanned by the matrices denoted by the letters,
 with the (\ref{letters}) identification, given by:
\bea
\label{z2sectors}
M_{0}: \quad I,~X; && M_{1}: \quad Y, ~A.
\eea
In the ${\mathbb Z}_2\times{\mathbb Z}_2$-grading the $M_{00}, M_{01}, M_{10}, M_{11}$ sectors can be spanned by the matrices denoted by the $2$-character words
\bea 
M_{00}: \quad II,~IX,~XI, ~XX; && M_{01}: \quad IA, ~IY, ~XA, ~XY;\nonumber\\
M_{10}: \quad AI,~AX,~YI, ~YX; && M_{11}: \quad AA, ~AY, ~YA, ~YY.
\eea
In Section {\bf 4} we showed that each one of the $7$ inequivalent ${\mathbb Z}_2$-graded superdivision algebras admits an alphabetic presentation in terms of equal-length words. Without loss of generality (up to similarity transformations) the even sector ${\mathbb D}_0^{[1]}$ can be expressed as
\bea\label{z2even}
&{\mathbb R}{\textrm{-series}}: ~ I;\qquad ~{\mathbb C}{\textrm{-series}}:  ~II,~IA;\qquad ~{\mathbb H}{\textrm{-series}}:  ~III, ~IIA,~IAX,~ IAY. &
\eea
The odd sectors  ${\mathbb D}_1^{[1]}$ are presented in table (\ref{z2real}) (for the ${\mathbb R}$-series), table
(\ref{z2complex})  (for the ${\mathbb C}$-series), table (\ref{z2quat}) (for the ${\mathbb H}$-series).\par
The alphabetic presentation is extended to the \zzg superdivision algebras by taking into account that:
\\
{\it i}) without loss of generality (up to similarity transformations) the even sector ${\mathbb D}_{00}^{[2]}$ can be expressed as
\bea
&{\mathbb R}{\textrm{-series}}: ~ II;\quad ~{\mathbb C}{\textrm{-series}}:  ~III,~IIA;\quad ~{\mathbb H}{\textrm{-series}}:  ~IIII, ~IIIA,~IIAX,~ IIAY;&
\eea
{\it ii}) each one of the three subalgebras ${\mathbb S}_{10}, {\mathbb S}_{01}, {\mathbb S}_{11}\subset {\mathbb D}^{[2]}$, given by the direct sums
\bea\label{ssubalg}
&{\mathbb S}_{01}:= {\mathbb D}_{00}^{[2]}\oplus{\mathbb D}_{01}^{[2]},\qquad
{\mathbb S}_{10}:= {\mathbb D}_{00}^{[2]}\oplus{\mathbb D}_{10}^{[2]},\qquad
{\mathbb S}_{11}:= {\mathbb D}_{00}^{[2]}\oplus{\mathbb D}_{11}^{[2]},&
\eea 
is isomorphic to one (of the seven) ${\mathbb Z}_2$-graded superdivision algebra;\\
{\it iii}) the alphabetic presentation can be assumed for ${\mathbb S}_{01}$ and, since the second  ${\mathbb Z}_2$ grading is independent from the first one, ${\mathbb S}_{10}$. The closure under multiplication for any $g\in {\mathbb D}_{01}^{[2]},~ g'\in
{\mathbb D}_{10}^{[2]}$ implies that $gg'\in {\mathbb D}_{11}^{[2]}$ is alphabetically presented.
\par
As discussed in Section {\bf 5}, a \zzg superdivision algebra can be associated with its ${\mathbb Z}_2$-graded superdivision algebra projections ${\mathbb S}_{01},~{\mathbb S}_{10},~{\mathbb S}_{11}$.

~\par
~\par
\par {\Large{\bf Acknowledgments}}
{}~\par{}~\par

 The work was supported by CNPq (PQ grant 308846/2021-4).

~\par
~\par
\par {\Large{\bf Declaration of Conflicting Interest}}
{}~\par{}~\par

 The authors declare that there is no conflict of interest.

~\par
~\par
\par {\Large{\bf Data availability statement}}
{}~\par{}~\par

 The authors can confirm that all relevant data are included in the article.

\end{document}